\documentclass[pre,superscriptaddress,showpacs,floatfix,aps,preprint]{revtex4}

\usepackage[utf8]{inputenc}
\usepackage{graphicx}
\usepackage{amsmath}
\usepackage{amsfonts}
\usepackage{bm}
\usepackage{color}
\usepackage{epstopdf}
\usepackage{natbib}
\usepackage{hyperref}

\newcommand{\cch}[1]{\left[#1\right]}

\newcommand{\prt}[1]{\left(#1\right)}
\newcommand{\aver}[1]{\left\langle #1 \right\rangle}

\begin{document}
\title{Temperature of maximum density and excess properties of short-chain alcohol
aqueous solutions: A simplified model simulation study.}

\author{A. P. Furlan}
\affiliation{Instituto de Física, Univeridade Federal do Rio Grande
  do Sul, Caixa Postal 15051, 91501-570, Porto Alegre, Rio Grande
  do Sul, Brazil.}

\author{E. Lomba}
\affiliation{Instituto de Química Física Rocasolano,CSIC, Serrano 119,
  E-28006 Madrid, Spain}

\author{M. C. B. Barbosa}
\affiliation{Instituto de Física, Univeridade Federal do Rio Grande
  do Sul, Caixa Postal 15051, 91501-570, Porto Alegre, Rio Grande
  do Sul, Brazil.}

\begin{abstract}
  We perform an extensive computational study of  binary mixtures
  of water and short-chain alcohols resorting to two-scale
  potential models to account for the singularities of hydrogen bonded
  liquids. Water molecules are represented by a well studied core
  softened potential which is known to qualitatively account for a
  large number of water's characteristic anomalies. Along the same lines,
  alcohol molecules are idealized by dimers in which the hydroxyl
  groups interact with each other and with water with a core softened
  potential as well. Interactions involving non-polar groups are all
  deemed purely repulsive. We find that the qualitative behavior of excess
  properties (excess volume, enthalpy and constant pressure heat
  capacity) agrees with that found experimentally for alcohols such as
  t-butanol in water. Moreover, we observe that our simple solute
  under certain conditions acts as an   ``structure-maker'', in the
  sense that the temperature of maximum 
  density of the bulk water model increases as the solute is
  added, i.e. the anomalous behavior of the solvent is enhanced by the
  solute. 
\end{abstract}
\maketitle


\section{Introduction}

Processes involving mixtures of water and a variety of organic compounds are present in a
huge diversity of phenomena. In most cases, effects of hydrogen
bonding and hydrophobicity are the key elements in determining the
behavior of such mixtures. These range  from the simplest case of
diluted short chain alcohols, to the substantially more involved situations
of biomolecules (e.g. proteins) in solution. The former have attracted
special attention from the technological standpoint due to their relevance in the 
bioethanol industry~\cite{Ja16,Di13} as well as in pharmaceutical and
cosmetic industries, being some of the preferred solvents for a wide
range of solutes with varying degrees of polarity. Moreover, from a
fundamental perspective the study of dilute short chain alcohol
solutions is of utmost importance, being the simplest systems that
illustrate the interplay of hydrogen bonding and hydrophobicity in
amphiphilic substances. In particular, their thermodynamics is known
to exhibit quite a few characteristic features, such as the presence of maxima in the
excess specific heat~\cite{Fr66,Be80}, minima in the excess
volume~\cite{Fr66,Pa85} and negative excess entropy~\cite{Dx02}. Some
of the anomalies found in these systems are in close connection with
the  more than seventy anomalies present in water,
among them, the presence of a density maximum at 3.98 C and 1 bar in liquid water~\cite{Ke75}.

The anomalous behavior
of water in the fluid phase has been explained in terms of the competition between
a low density structure dominated by the presence of hydrogen bonds
and exhibiting essentially tetrahedral ordering, and a high density
one, with higher coordination and a much lower degree of hydrogen
bonding. For temperatures above the the temperature of maximum density
(TMD), the high density structure dominates, and the system expands
upon heating, whereas below the TMD the loosely coordinated structure
is the preferred one and the system contracts upon heating, due to the
breakup of the hydrogen bond network. 

The way in which solutes modify the anomalous properties of water
is not yet completely understood. In this respect, concerning  the density anomaly,
solutes can be classified into two groups, namely ``structure-makers''
(as they increase the TMD when added to water) or ``structure-breakers'' (decrease the
TMD)~\cite{Da68,He69}. It has been found that solutes whose molecules  do not
join the hydrogen bond network, such as 
electrolytes~\cite{Darnell1968}, room temperature ionic liquids~\cite{Ta15}, or polar substances without H-bond active groups~\cite{Wa62}
(e.g. acetone, acetonytrile, tetrahydrofurane, among others) induce a
decrease  of the TMD. They can all be cast into the group of ``structure-breakers''. In contrast, dilute solutions of hydrogen bond forming
substances with relatively small non-polar tails, such as  short-chain
alcohols~\cite{Wada1962} and some amines~\cite{Franks1967}, exhibit an
increase of the TMD with respect to pure water. This substances are
thought to enhance the structuring of the tetrahedral low density
phase of water, and thus are ``structure-makers''. The change in the TMD
is measured in terms of,
$\Delta T_{MD}=T_{MD}\prt{x_2}-T_{MD}(x_2=0)$ where $T_{MD}\prt{x_2}$
represent the temperature of maximum density of the solution at a given
molar fraction of solute, $x_2$, and $T_{MD}(x_2=0)$ obviously refers to the TMD
of pure water. Typically, for a structure-maker $\Delta T_{MD}>0$ until a given solute mole 
fraction, for which a maximum is reached, and then it decreases up to
a certain concentration where the solution no longer presents a
maximum density in terms of temperature. From the work of Wada and
Umeda~\cite{Wada1962} it was found that the largest increase in the TMD at atmospheric
pressure occurs for
t-butanol at  $x_2\simeq 0.0043$ with a $\Delta T_{MD}=0.41$ K.

In close connection with the anomalies found in dilute
hydrogen-bonding water solutions,  the excess
mixture properties are also known to exhibit certain
singularities. Thus, for instance, in the case of small linear chain
alcohols (methanol~\cite{Ma00,To86} and ethanol~\cite{Ma00,Ot86}) and
alkylamines~\cite{Mu07}) the excess enthalpy is negative, whereas for
somewhat larger non-polar tails  (propanol and butanol), this same property
changes its concavity and assumes positive values.  Ionic liquids
behave differently depending on the degree of hydrophobicity of their
non-polar tails: less hydrophobic
ionic liquids have a negative excess mixture enthalpy, whereas
hydrophobic ionic liquids display positive excess enthalpies of mixing
with  a maximum. The excess mixture volume at ambient conditions is negative and
exhibits a minimum for alcohols~\cite{Pa85,Be08,Sh16,Ot93} and
alkylamines~\cite{Ya94,St14}. Similarly to the excess enthalpy, the excess
volume in ionic liquid aqueous solutions depends on the degree hydrophobicity of the
solute~\cite{Ha03}. The excess specific heat displays a peculiar
behavior for small alcohol concentrations, e.g.
in water-methanol mixtures
a maximum occurs at solute concentration $x_{Me OH}=0.16$~\cite{Be80}, and
for t-butanol~\cite{Subramanian2013} for $x_{t-BuOH}\approx 0.08$.

The first attempt to explain the excess thermodynamics of amphiphilic aqueous
solutions dates back to 1945, with the pioneering work of
Franks{ \it et.al.}~\cite{Fr45}. They establish a connection between
structure and the thermodynamics of mixtures and formulated the
``iceberg theory''. According to this interpretation, the presence of the
composition-dependent anomalies can be ascribed to the formation of a low
entropy cage of water with strong hydrogen-bonds around the alcohol
molecules, which in this case would increase the structural ordering
of bulk water. These ideas seem to have been confirmed by a series of
experimental X-ray  diffraction studies, in
most cases complemented by molecular
simulations~\cite{Nagasaka2014,Galicia-Andres2015}. In these studies it
was found that adding methanol to water, enhances the local three
dimensional network of water in the vicinity of the methanol
molecules, which explains the decrease in the entropy of mixing and
the considerable increase in the heat capacity at low alcohol
concentrations. 

In an attempt to provide a more quantitative analysis,  Chatterjee, Ashbaugh and
Debenedetti~\cite{Chatterjee2005} resorted to
a simple statistical mechanical model, and ascribed the increase in the TMD to
the hydrophobicity of the solute (the non-polar tail). Within their
approach, the dispersive solute-solvent 
interactions are thought to be responsible for the decrease of the density
anomaly to lower temperatures. Their picture, however, downplays the
role that hydrogen bonding must necessarily play in the phenomenology
of alcohol (or alkylamine)-water mixtures, and the essentially
different behavior of other polar solutes, such as acetonytrile or
acetone~\cite{Wa62}. Somewhat more sophisticated models have been
developed in which the alcohol is represented using site-site molecular
models (in the simplest case of methanol, a dimer), and the hydrogen
bonding interactions are modeled using a two-scale potential, both of
core softened type~\cite{Hu14,Hus2014b,Hu15,Munao2015} and a Jagla
type~\cite{Ja01} ramp potential~\cite{Su12}. These models give rise to
features such as liquid-liquid equilibria~\cite{Hus2014b} or the
presence of a TMD curve in methanol, which have not been confirmed
experimentally. On the other hand, these models can be fitted to reproduce
qualitatively the behavior of the excess mixture
properties~\cite{Su12}. In none of these cases, the influence of the
alcohol on the change of the TMD of water has been reproduced.

More sophisticated models can be tackled resorting to  computer
simulation. In these cases, the mixture has been modeled using all-atom site-site
interaction potentials, such as TIP4P/2005 for water~\cite{abascal05b}
or OPLS for alcohols~\cite{Jorgensen86}. Concerning structural features
of alcohol-water mixtures,
Allison{\it et.al.}~\cite{Al05} showed using Molecular Dynamics that the number of
hydrogen-bonds decreases and the water molecules become distributed in
rings and clusters as alcohol concentration increases, in accordance
with experimental results~\cite{Dx02}. 
Laaksonen{\it et.al.}~\cite{La97} simulation results indicated that
the system is highly ordered
around the hydroxyl groups, and  methanol molecules are solvated
by water molecules, in accordance with the assumptions of Iceberg
theory~\cite{Fr45,Fr66}. Also using molecular dynamics simulations,
Bako and coworkers showed that the despite decreasing  number of hydrogen
bonds in the mixture, the tetrahedral structure of water is
preserved~\cite{Ba08}. A recent mixture model for methanol-water developed by
González-Salgado and coworkers~\cite{Gonzalez-Salgado2016} has shown to
be able to reproduce quantitatively the excess thermodynamic
properties, but does not account for the ``structure-maker'' character
of methanol molecules at high dilution, and in fact the decrease in
the TMD with concentration predicted by the model is practically one
order of magnitude larger than the experimental. 

It is then clear that there is much room for improvement in our
knowledge of the dilute solutions of hydrogen bonding substances in
water. In this work, we aim at obtaining further insight using
continuous site-site two-scale potential models, which are simple
enough to discriminate the different effects than enter the structural and
thermodynamic behavior of the model, but at the same time are able to
reflect the anomalous features of water-like systems. To that aim,  we
have here studied the excess thermodynamics and the
density anomaly of a mixture of water and
amphiphilic dimers, in which water (solvent) is represented by a
spherically symmetric two length scale
potential~\cite{Si10}, and the alcohol molecules (dimers)  are modeled by a repulsive
R-site  and  an OH-site which interacts  with other OH sites and with
water by means of a two length scale potential. This system will be
studied by means of extensive Molecular Dynamics calculations in 
various ensembles, with different two sizes of the apolar site of the
alcohol-like molecule,  a methanol-like model
(homonuclear) and a tert-butanol-like one (heteronuclear), in which the R-site is
substantially larger than the OH-site. 

The remaining of the paper is organized as follows. In Section~\ref{sec:model}
we present our models for water and alcohol molecules.
 In Section~\ref{sec:sim} relevant technical details of the
 simulations are presented. Next, in Section 
\ref{sec:results} the results for our hetero- and homonuclear alcohol
models in solution are introduced, both concerning excess properties
and  influence on the TMD of water. A brief summary and a presentation
of our main conclusions and future prospects close this article
in Section~\ref{sec:concl}.

\section{The Model}\label{sec:model}
\begin{figure}[!htb]
  \centering
    \includegraphics[scale=0.7]{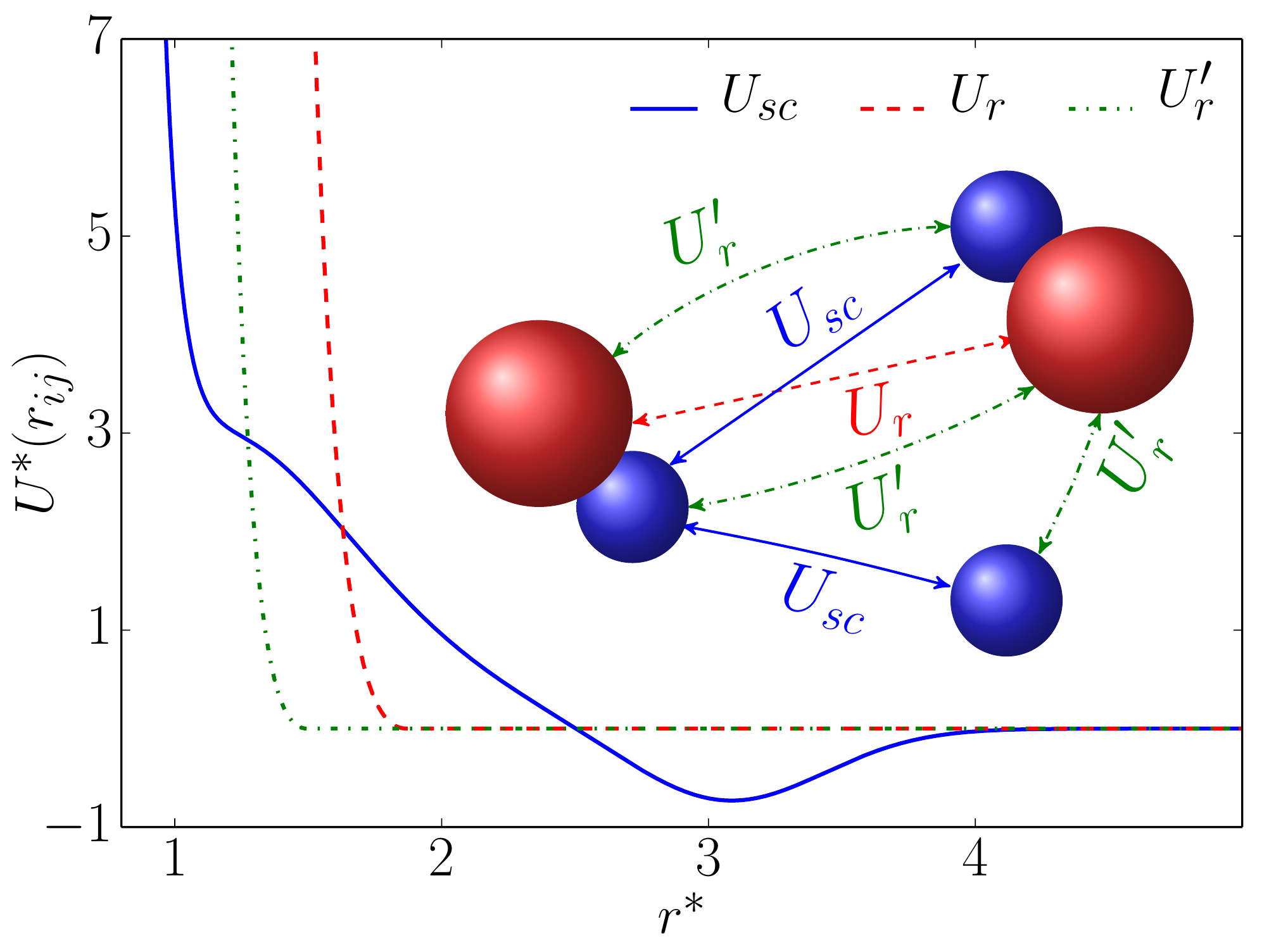}
  \caption{Interaction potential $versus$ distance. The solid blue
    line represents the softened-core interaction potential $U_{sc}$
    (equation~\ref{eq:model}) between OH-OH, OH-water and water-water
    sites. The dashed red line and 
    green dot-dashed line represent the R-R, R-OH and R-water
    repulsive interactions.}
  \label{fig:potentials}
\end{figure}

As mentioned above, we will have spherical particles representing
water-like molecules, together with an amphiphilic solute with a purely
repulsive site accounting for the apolar tail, R, in addition to an OH
site, characterized by  OH-OH and OH-water interactions with two
length scales~\cite{Si10}. A short range repulsion accounts for the
high density liquid phase, and a much longer range repulsion and
attraction attempts to roughly model the more open structures due to
hydrogen bonding. To make matters simpler, we will use the same  
softened-core
potential both for water-water, OH-OH and OH-water interactions,  defined by,
\begin{equation}
  \label{eq:model}
  U_{sc}(r_{ij})=4\epsilon_{sc}\cch{\prt{\dfrac{\sigma}{r_{ij}}}^{12}-
    \prt{\dfrac{\sigma}{r_{ij}}}^6} + \sum_{\ell=0}^1u_\ell\epsilon_{sc}\exp
  \cch{-\dfrac{1}{c_\ell^2}\prt{\dfrac{r_{ij}-r_\ell}{\sigma}}^2}.
\end{equation}
Here, $r_{ij}$ represents the separation between sites $i$ and
$j$. The first term on the r.h.s. 
 of Equation~(\ref{eq:model}), is the standard 12-6 Lennard-Jones
(LJ) potential~\cite{Aln89}, whereas the second term is the summation of two
 Gaussians, centered at $r_0=0.7\sigma$ and $r_1=3\sigma$, with depths
$u_o=5$ and $u_1=-0.75$ and widths $c_0=1$ and $c_1=0.5$
respectively. The potential of Eq.~(\ref{eq:model}),
displays two different length scales, an attractive scale at
$r \approx  3\sigma$ and a repulsive shoulder at
$r  \approx \sigma$. Of the many possible choices of two-scale
potentials, ours has been motivated by its ability to account
for many of the anomalous features of fluid
water~\cite{Si10,Kel67,Ang76}, displaying the characteristic
cascade ordering of anomalies~\cite{Egt01}. For the parameters chosen
in this work, the model is known to display a density anomaly with a
TMD curve in the supercritical region~\cite{Si10}. The attractive well
that can be seen in Figure~\ref{fig:potentials} is not sufficient to
place the anomalous region within the stable liquid phase, in contrast
with the situation in real water. Despite these limitations, as
already mentioned, this model potential is an excellent candidate to
reproduce water anomalies~\cite{Si10}.

The non-polar site-site interactions (R-R, R-OH, and R-water) are
represented by a 
purely repulsive Weeks-Chandler-Andersen potential (WCA)~\cite{Wca54}
of the form
\begin{equation}
  U_{r}(r_{ij})=\left\{\begin{array}{lcl}
  U_{LJ}(r_{ij})-U_{LJ}(r_{c}) & {\rm if} &  r\leq r_m \\
  0                        & {\rm if} & r > r_m
  \end{array}\right.
  \label{eq:wca}
\end{equation}
where, $U_{LJ}(r)$ is the standard 12-6 LJ potential with parameters ($\epsilon_r,\sigma_r$), and $U_{LJ}(r_m)$
is the LJ potential computed at cutoff distance given by
the position of the minimum of the LJ interaction, $r_m=2^{1/6}\sigma_r$.

In what follows we have used as unit length,
$\sigma$=$\sigma_{ww}=\sigma_{w-OH}=\sigma_{OH-OH}$, and as energy
unit, $\epsilon_{sc}$. Reduced pressure and temperature are defined as
$P^*=P\sigma^3/\epsilon_{sc}$ and $T^*=k_BT/\epsilon_{sc}$, where
$k_B$ is Boltzmann's constant. The simulation time step is given in reduced
units of $\tau = \sigma\sqrt{m/\epsilon_{sc}}$, where $m$ is one
of the particle masses. Since here we are not interested in dynamic
properties, we have considered all particle masses identical. 

As mentioned, we have considered an heteronuclear model, in which
$\sigma_r/\sigma=5/3$ (a rough model for tert-butanol), and a
homonuclear model in which $\sigma_r/\sigma=1$. 

The energy parameter of the repulsive interaction  was set to
$\epsilon_{r}/\epsilon_{sc}=1.21$, and the dimer bond length to
$d_{R-OH}=0.48\sigma$. This choice of parameters was to some
extent inspired by the OPLS force field widely used to simulate
alcohols~\cite{Jorgensen86}.  Cross interaction parameters were
computed using Lorentz-Berthelot mixing rules~\cite{Aln89}. A
graphical representation of our molecular models and the corresponding
 interactions is depicted in Figure~\ref{fig:potentials}.

\section{Simulation details}\label{sec:sim}

Using the LAMMPS package~\cite{Plt95},  we have performed MD simulations
for a system with a number of particles ranging from 2000 to 4000 for
various compositions. 

The simulations  were performed in the $NPT$ ensemble with a Nosée-Hoover
thermostat and barostat~\cite{Nse84,Hvr85} and particles were placed in a cubic box with
standard periodic boundary condition.  The dimer bonds were kept fixed
using a  SHAKE algorithm~\cite{Rkt77}, with a tolerance
factor of $10^{-5}$. Since the system can undergo a demixing
transition, we have systematically checked that the thermodynamic
conditions under consideration were away from instability by inspection of
the small wave vector behavior of the concentration-concentration
structure factor~\cite{HansenBook2nd,Bores2015a}. For our mixture this
quantity is defined by
\begin{equation}
S_{cc}(Q)=x_{ROH}^{2}S_{ww}(Q)+x_{w}^{2}S_{ROH-ROH}(Q)-2x_{ROH}x_{w}S_{w-ROH}(Q),\label{scc}
\end{equation}
where $x_w$ and $x_{ROH}=1-x_w$ are the mole fractions of water and
alcohol respectively. For the partial structure factors,  we have  approximated
$S_{ROH-ROH}=S_{RR}$ and $S_{w-ROH}=S_{wR}$, i.e. we have neglected the
contribution of the OH-sites of the dimer. In the study of demixing,
this approximation is harmless, since the positions of R and OH sites
within the same molecule are obviously tightly bound. The
site-site structure factors are numerically determined 
from the spatial configurations generated during additional NVT
simulation runs (in order to keep the box size constant for the
binning procedure in Q-space) using standard procedures~\cite{Bores2015a}.

 The signature of concentration
fluctuations associated with demixing is typically a low-Q diverging
concentration-concentration structure factor. By monitoring this
quantity along our simulations we have ruled out the presence of
inhomogeneities due to demixing. 

Our simulations started from a compositionally disordered mixture of
ROH and water particles, which was equilibrated at the chosen pressure
and temperature for 1$\times 10^7$ steps in the NPT-ensemble. Production
runs were  8$\times 10^7$ step long. The time step was set to $5\times
10^{-6}\tau$ in reduced units. 

\section{Results}\label{sec:results}
In what follows we will present our results both for the hetero- and
homonuclear ROH models in a solution of our water-like fluid, first
focusing on the ROH influence on the temperature of the maximum density
curve of water (which was already determined in
Ref.~\onlinecite{Si10}). We will analyze the influence of the
alkyl-group size on the changes of the TMD, comparing the results of
our heteronuclear and homonuclear models. Finally, we will analyze the behavior of
the excess thermodynamics of the mixture just for the heteronuclear model. 

\subsection{The temperature of maximum density (TMD)}

\begin{figure}[!htb]
  \includegraphics[clip,scale=.4]{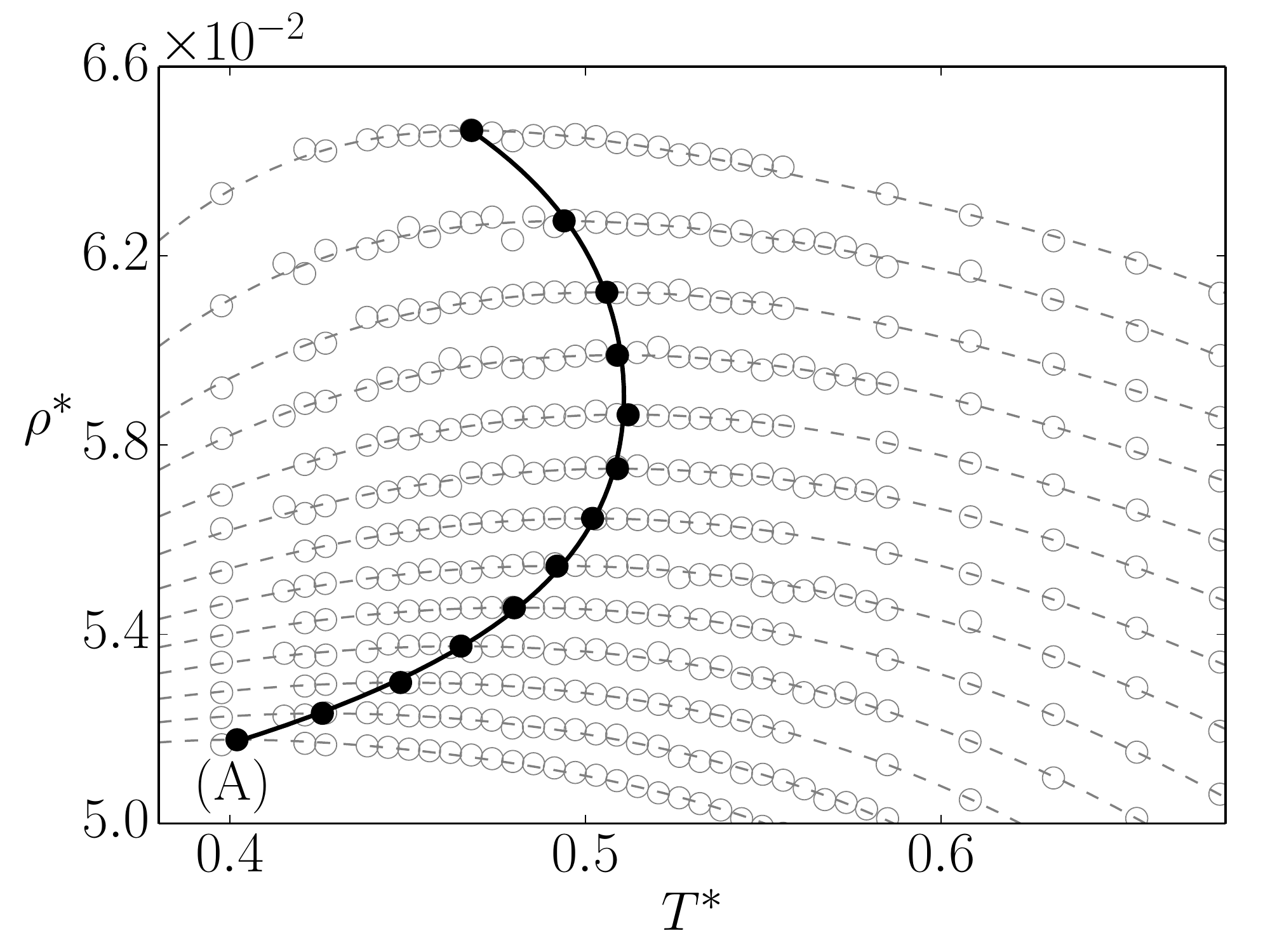}
  \includegraphics[clip,scale=.4]{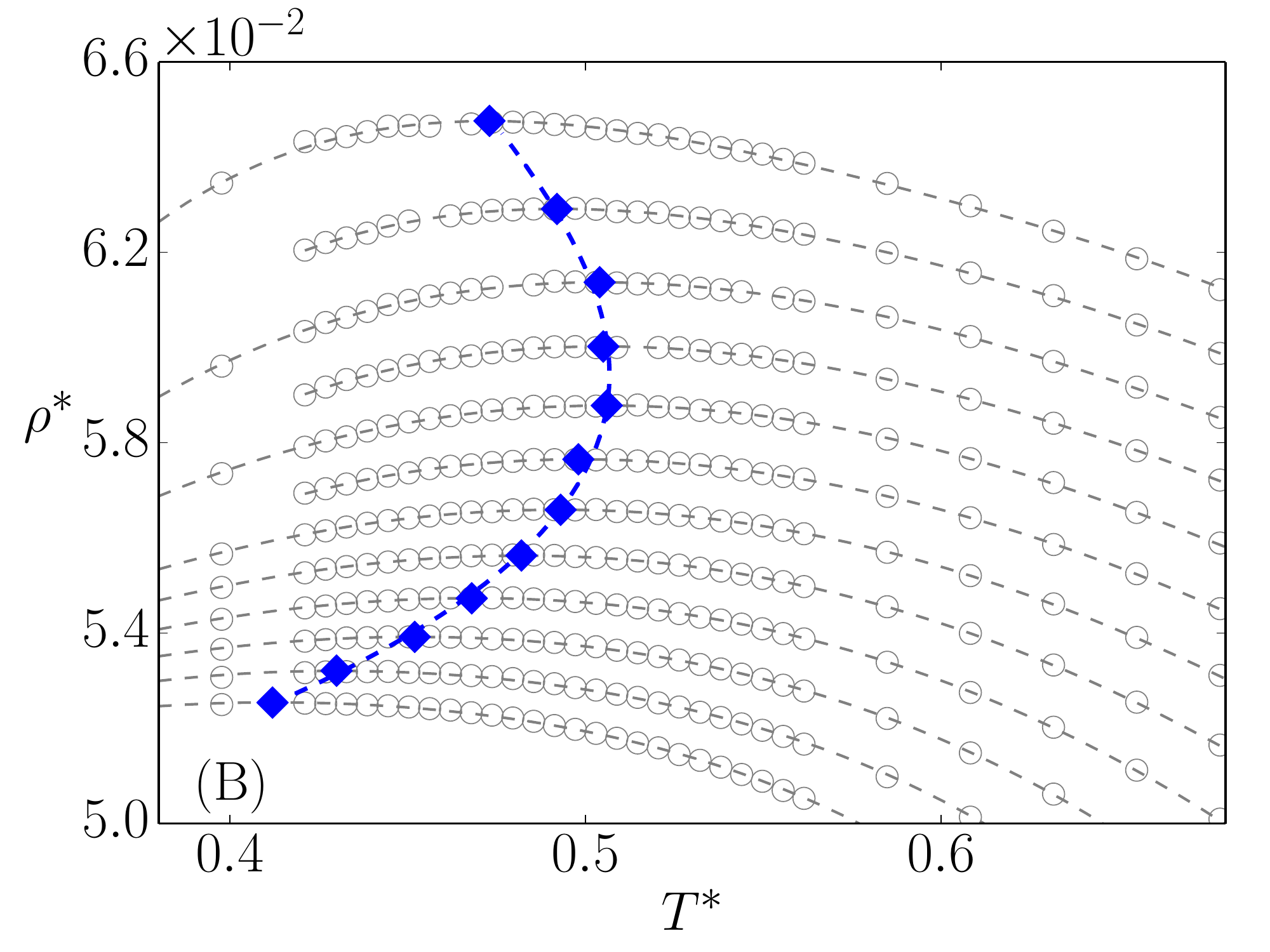}

  \includegraphics[clip,scale=.4]{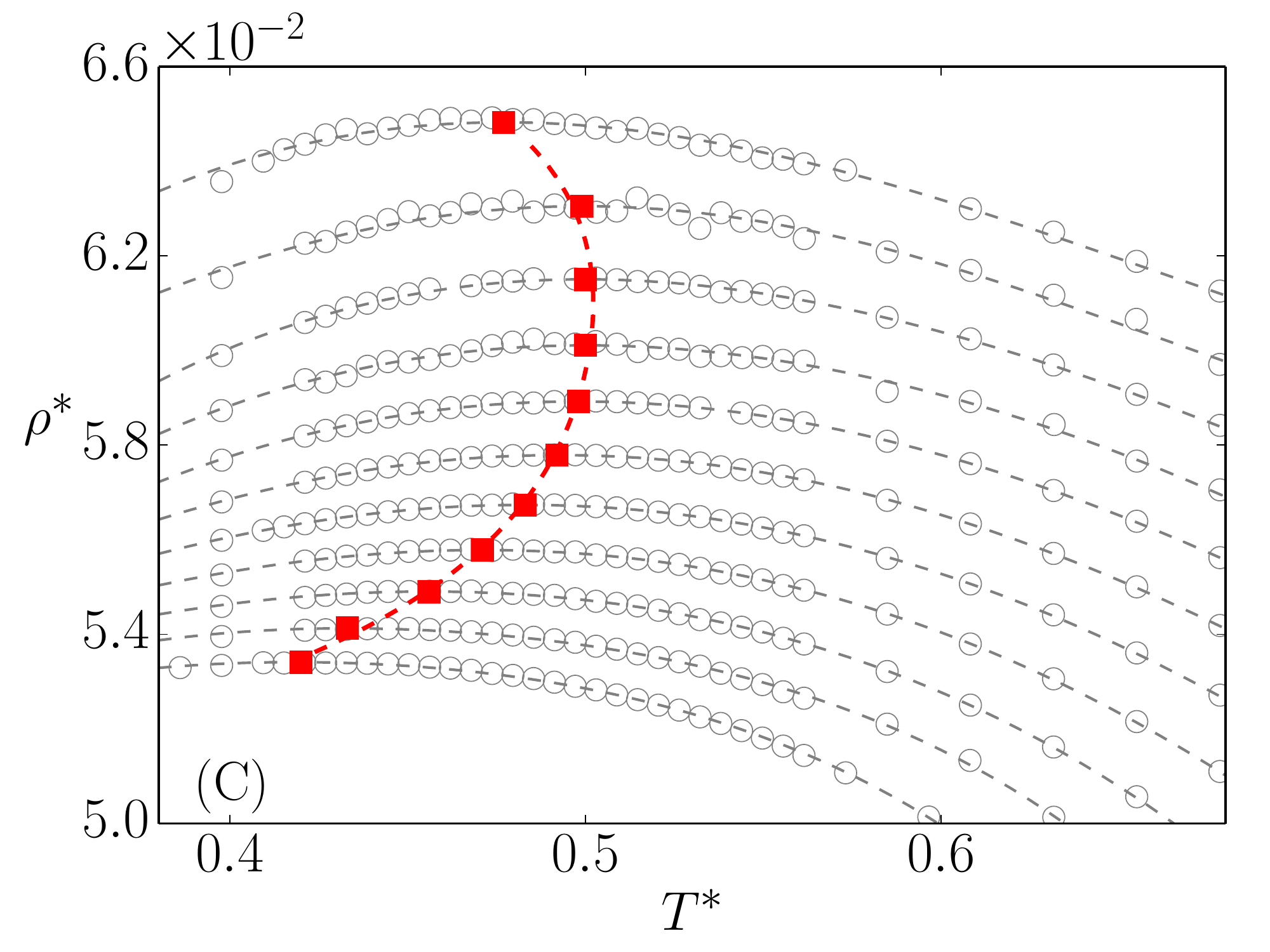}
  \includegraphics[clip,scale=.4]{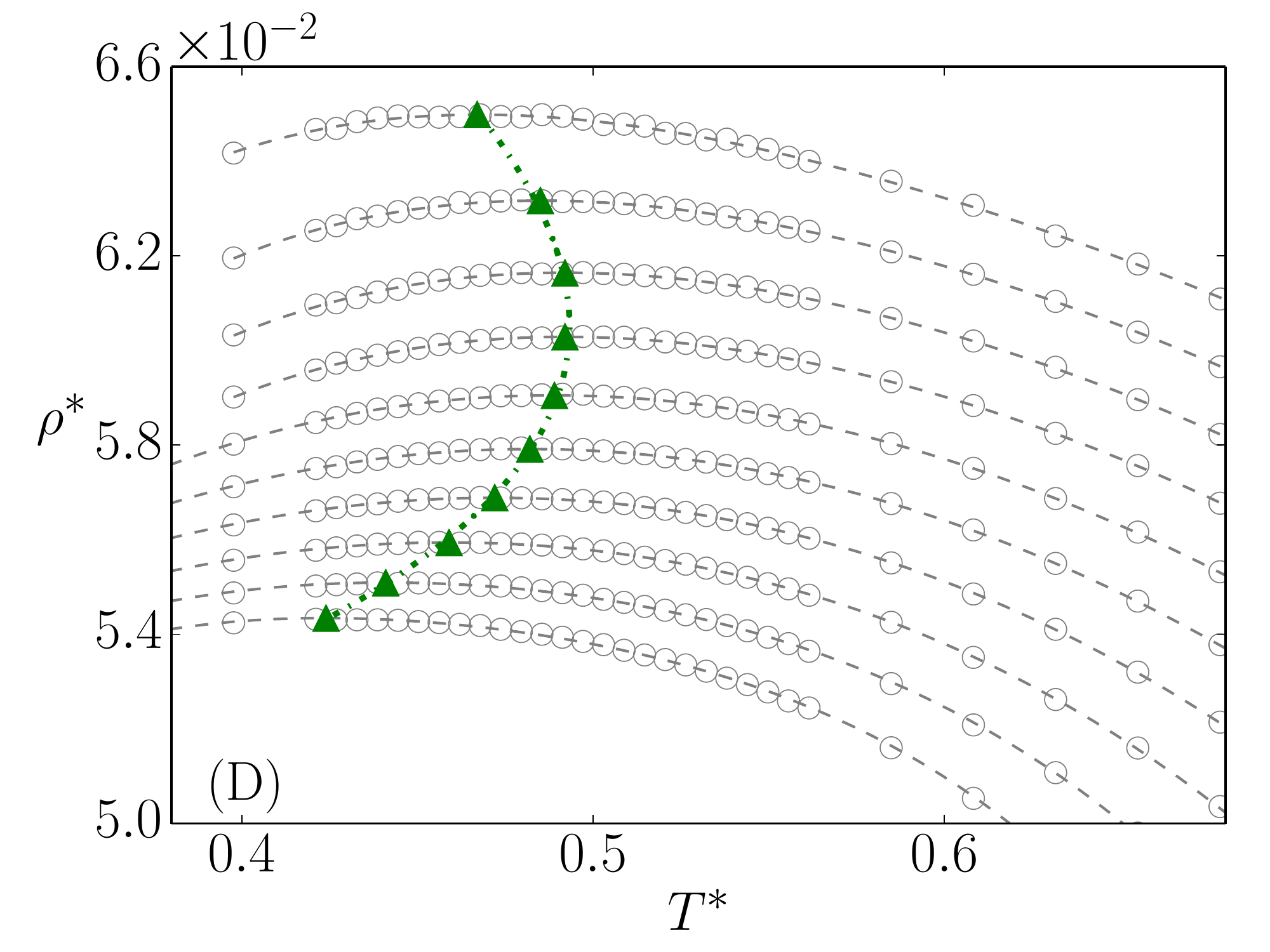}
  \caption{Temperature dependence of the density for various solute
    compositions along isobars with increasing pressure from bottom to
    top ($P^*=2.3,\ldots,27.6$). Open circles correspond to 
    simulation data and a dotted line denotes a polynomial
    fit. (A) $x_{ROH}=0.0$ (pure water), (B) $x_{ROH}=0.01$, (C)
    $x_{ROH}=0.02$ and (D) $x_{ROH}=0.03$. The TMD is represented by filled
   symbols, which are connected with a dashed curve that correspond to
   a polynomial fit, to represent the TMD curve. Pressure increases
   from bottom to top.}
  \label{fig:std_rhot}
\end{figure}

The density anomaly in water and ROH aqueous solutions can be easily
detected representing the temperature dependence  of the density along
isobars.  This can be done studying a series of state
points along various isobars by means of NPT simulations. These results
are presented in the Figure~\ref{fig:std_rhot} for various ROH mole
fractions, namely $x_{ROH}=0.00,0.01,\ldots,0.04$, first for our
heteronuclear model. Note that the apparent low values of the reduced density
are due to the fact that densities are scaled with the inner core of
the potential. If scaled with the range of the second repulsive range
($\approx 2.5\sigma$), which is a more appropriate measure of the
molecular size, we will have reduced densities in line with what one
should expect for a liquid ($\rho^* \approx 0.5 \sim 0.9$).

 At a certain
concentration of ROH the TMD disappears, since our ROH model lacks a
density anomaly.  A relatively accurate numerical estimate of the TMDs 
was  obtained by a polynomial fit to the simulated
densities. These points (denoted by solid squares) are connected in the
Figure \ref{fig:std_rhot} with short-dashed lines, that constitute the
TMD curve in the $T-\rho$ plane. We observe that the region on the
left of the TMD points is characterized by the typical density
anomaly, namely a density increase upon heating. Note that for all
compositions the TMD increases with pressure, to reach and maximum and
then decreases. This decrease of the TMD with pressure corresponds to
the experimental behavior found in water~\cite{Pi2009}, and is the result of
the destructuring effect of pressure on the open structures (hydrogen
bonded network in the case of water) whose interplay with the high
density phase gives rise to the density anomaly. The increase of the
TMD with pressure at low pressures is not found experimentally, and it
is a consequence of the fact that in our model the TMD curve is placed
in the supercritical region. This feature is present even in
models  for which the TMD curve is in a low
density liquid region, such as the ramp fluid~\cite{Lomba2007}. 

\begin{figure}[!htb]
  \includegraphics[clip,scale=.6]{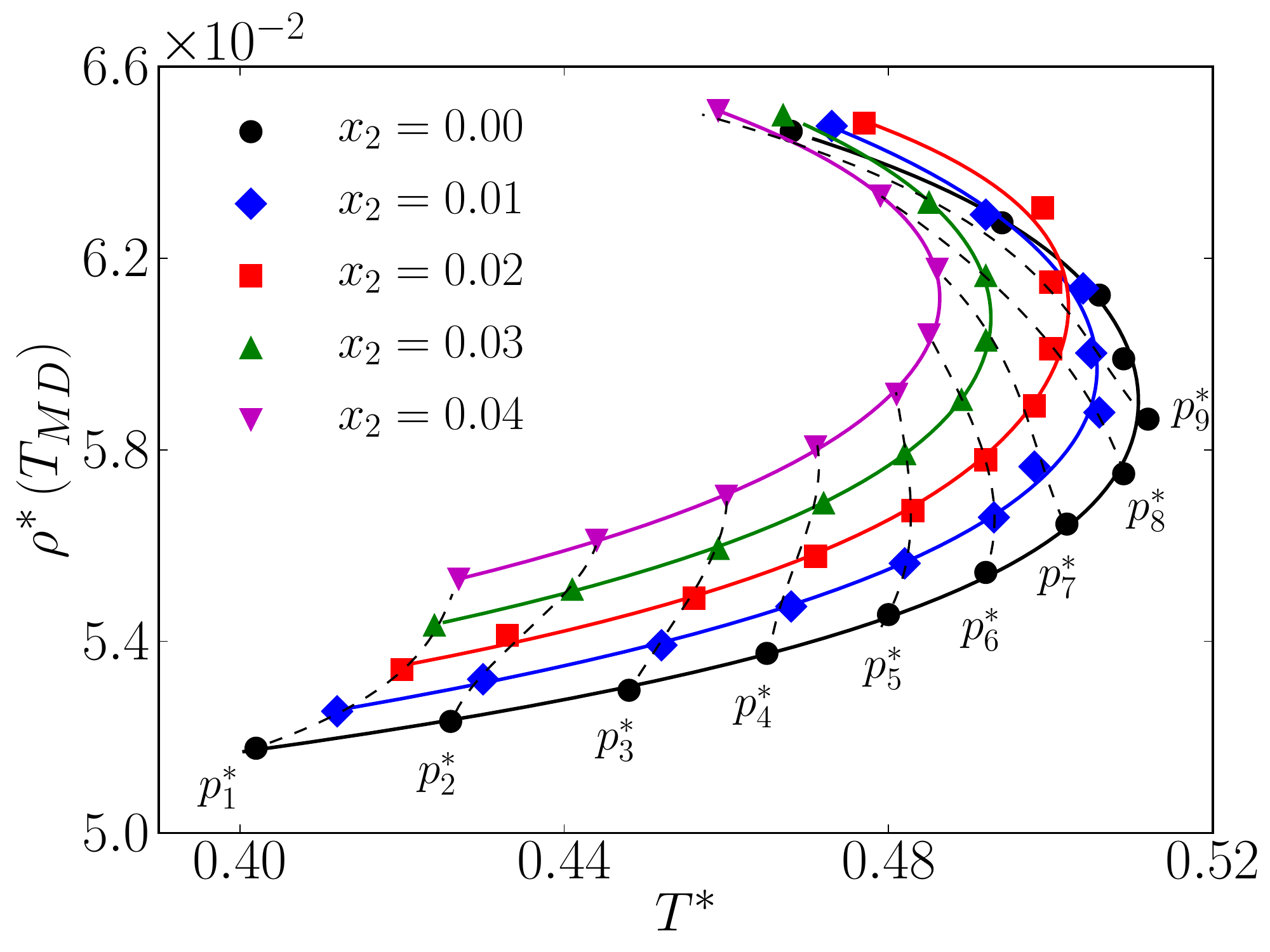}
  \caption{ Density values at the temperature of maximum density of
    the heteronuclear alcohol model, for various mole fractions,
    $x_{ROH}=0.00$ (black 
    solid lines and filled circles ), $x_{ROH}=0.002$ (red dashed and
    filled squares) and $x_{ROH}=0.003$ (green dot-dashed line and filled
    triangles). In all cases the points are simulation data, and lines
    correspond to polynomial fits. State points at the same pressure
    are connected with dotted lines.}
  \label{fig:fig_tmds_std}
\end{figure}

The various TMD curves for different mole fractions are represented in 
the Figure~\ref{fig:fig_tmds_std}. One readily appreciates that the
addition of alcohol reduces the density range and the temperature at
which the density anomaly is found, ultimately leading to its
disappearance. Points at equal pressure are connected by dashed
lines. 

\begin{figure}[!htb]
  \includegraphics[clip,scale=.6]{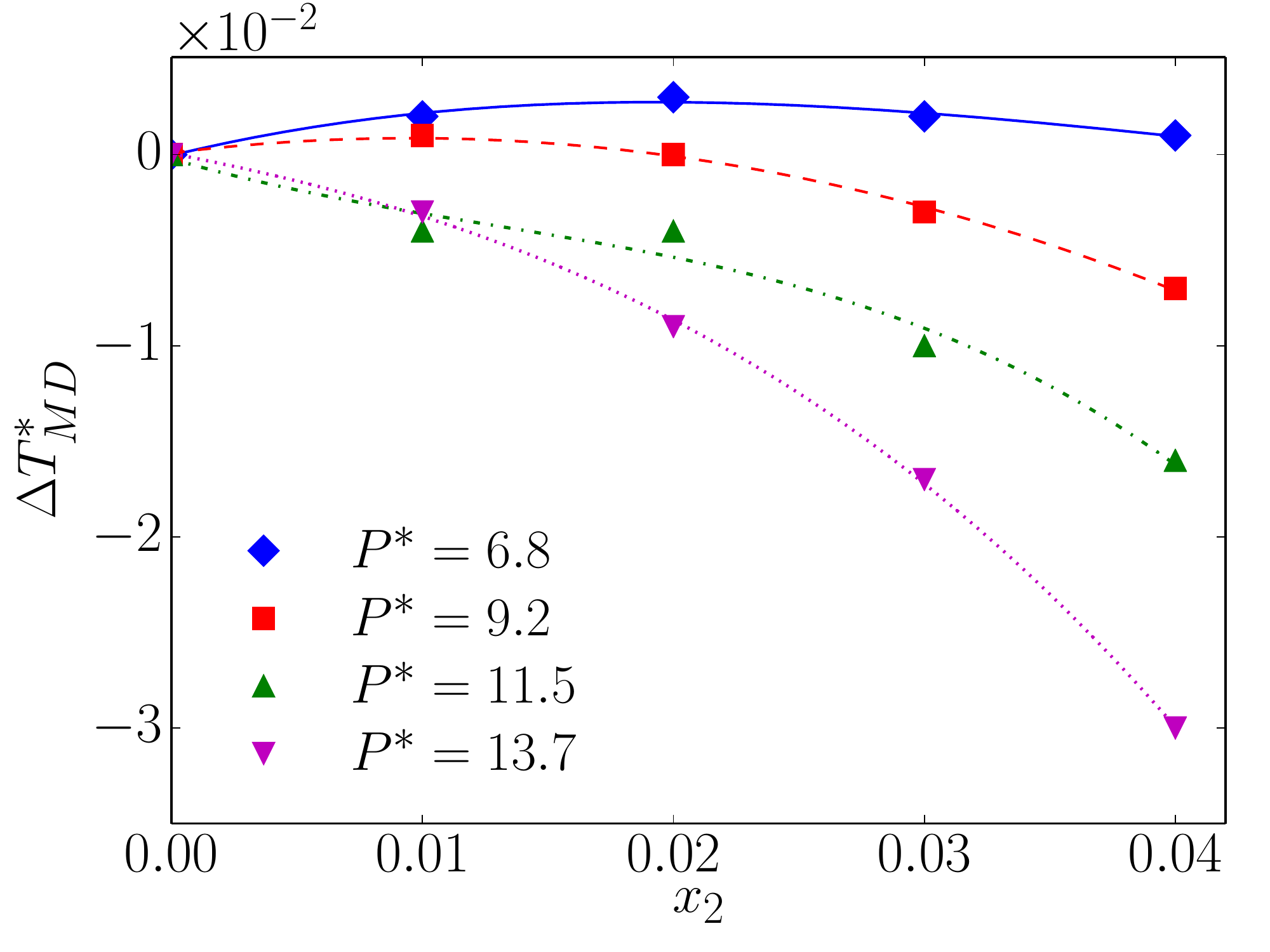}
  \caption{Change in the  temperature of maximum density with respect
    to the bulk solvent value vs. alcohol mole
    fraction for various pressures for the heteronuclear alcohol model.}
  \label{fig:tmds_P}
\end{figure}

The change in the TMD with respect of that of pure water
($\Delta T_{MD}(x_{ROH})= T_{MD}(x_{ROH})-T_{MD}(x_{ROH}=0))$ induced
by the presence of solute  is represented in
the Figure~\ref{fig:tmds_P} for various pressures.  For pressures below
$P^*\approx 10$, and up to a certain concentration, we observe that
our solute acts as a  
``structure-maker''. This means that the presence of solute molecules
enhances the anomalous behavior of water, 
by favoring the build up of open structures and hence increasing the TMD. At $P^*=9.2$, the curve
presents a maximum around $x_{ROH}=0.03$ and then decays, which is the
qualitative behavior of the TMD tert-butanol in
water~\cite{Wada1962}. We find that as pressure increases the change in
the TMD is lowered, and as a matter of fact for $P^* > 10$, $\Delta
T_{MD}(x_{ROH}) <0$, and the solute behaves as a
``structure-breaker'', reducing the range of anomalous behavior of
water. This is accordance with the fact that the 
increase of pressure tends to destroy the low density
structures that give rise to the density anomaly, therefore the
structuring effect of the solute decreases, to finally turn the
``structure-maker'' into a ``structure-breaker''. For sufficiently
high pressures our alcohol-like molecules behave like standard solutes
which tend to decrease the TMD~\cite{Wa62}, i.e. the effect of the
two-scale interaction stemming from the OH site is no longer apparent
for sufficiently high pressures. A parallel situation occurs with the effect of the
hydrogen bonds in water when pressure starts to break them.

\begin{figure}[htb]
  \includegraphics[clip,scale=.7]{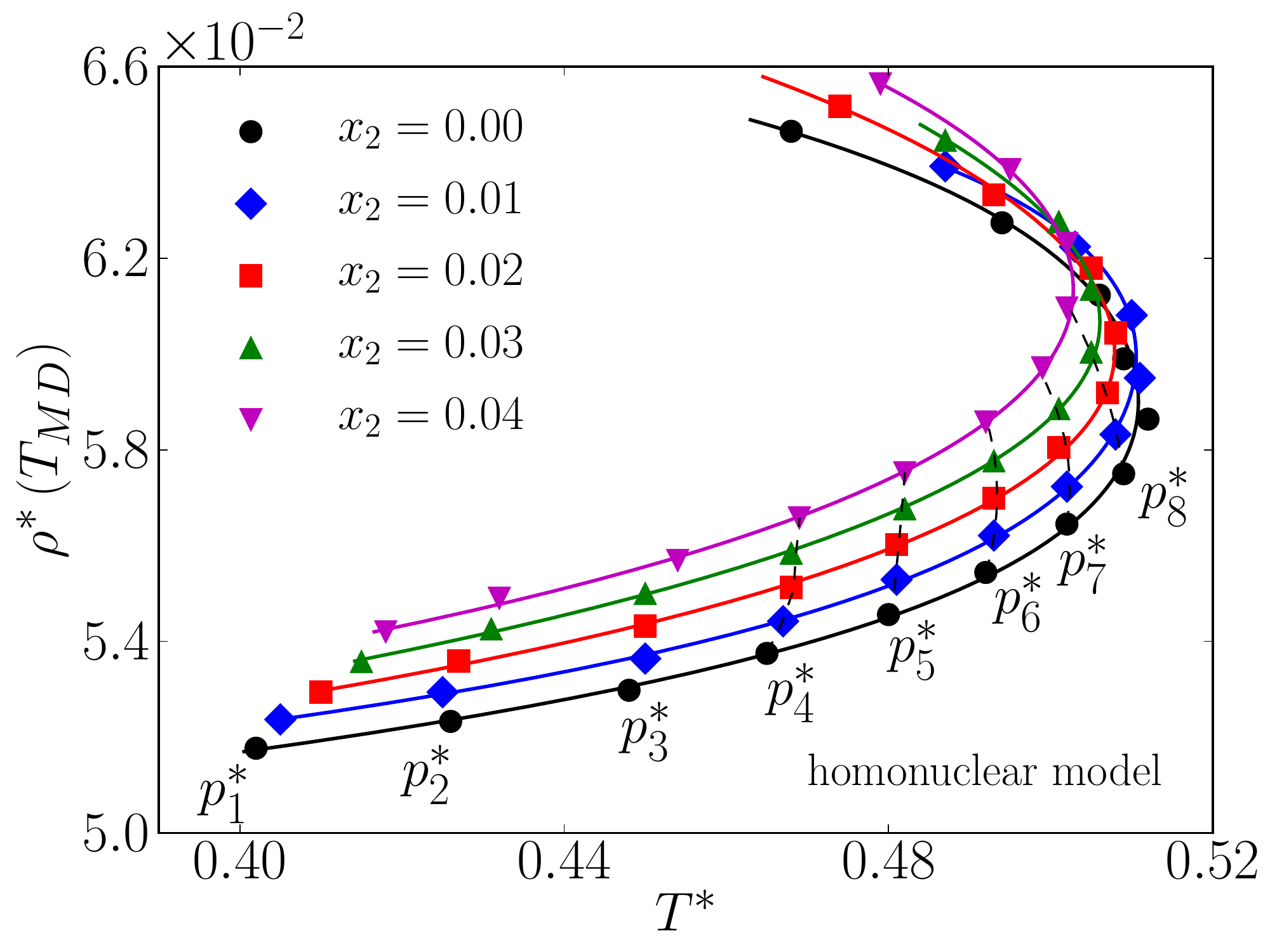}
\caption{Same as Figure \ref{fig:fig_tmds_std} for the homonuclear
  alcohol model}
\label{fig:homotmd}
\end{figure}

\begin{figure}[htb]
  \includegraphics[clip,scale=.7]{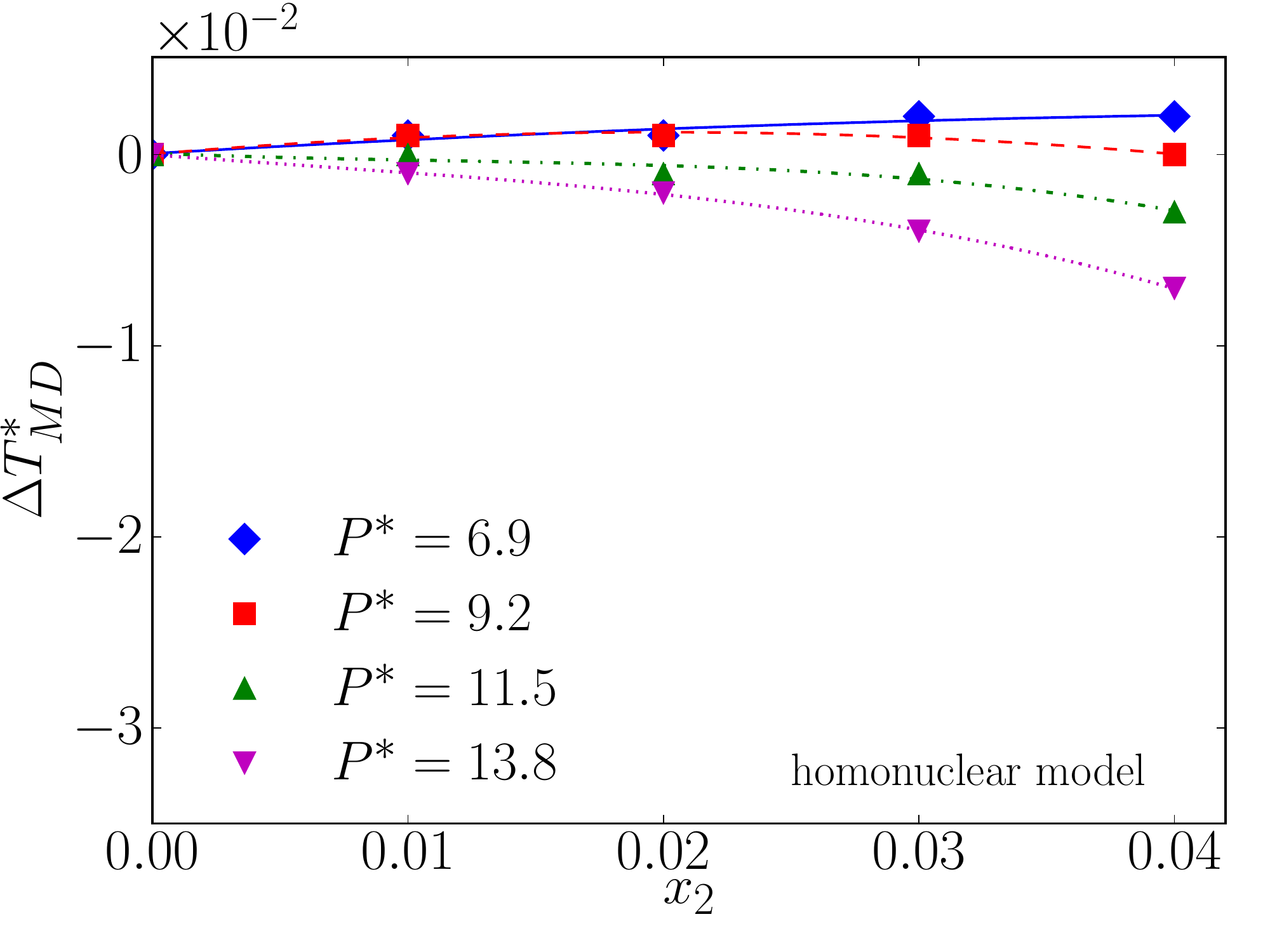}
\caption{Change in the  temperature of maximum density with respect
    to the bulk solvent value vs. alcohol mole
    fraction for various pressures for the homonuclear alcohol model.}
\label{fig:homodeltmd}
\end{figure}

Now in the Figure~\ref{fig:homotmd} we present the corresponding $\rho-T$
TMD curves for the homonuclear model of alcohol in solution. The first
effect one can observe is the shift of the TMD curves as a function of
solute concentration is minimized with respect to that observed in the
Figure~\ref{fig:fig_tmds_std} for the heteronuclear case. This is a
clear indication that the larger the size of the apolar tail of the
ROH, the more significant the effect of the solute on the TMD. 
The size dependency of the anomalous behavior is more clearly
illustrated in the Figure~\ref{fig:homodeltmd}, where change in the TMD,
$\Delta T_{MD}(x_{ROH})$ for the homonuclear model is represented as a
function of alcohol concentration, $x_{ROH}$ for various
pressures. Note that the same scale as in the Figure \ref{fig:tmds_P} is
used. Comparison of both figures shows that the increase in size of
the apolar tail of the alcohol increases the changes in the TMD. On
one hand, for pressures below $P^*\approx 10$ the maximum in $\Delta
T_{MD}(x_{ROH})$ (a characteristic of t-butanol and ethanol in
dilution~\cite{Wada1962}), practically disappears for the homonuclear
model. Interestingly, this model displays a behavior resembling that
of methanol~\cite{Wada1962}, for which the maximum is hardly visible. For
this values of the pressure, the ``structure-maker'' character of the
model alcohol is enhanced when the apolar chain is larger. This is in
agreement with the experimental data, and with the theoretical
predictions of Chatterjee et al.~\cite{Chatterjee2005} statistical
mechanical model for solutions of apolar solvents in water. Now, as
pressure increases above, $P^*\approx 10$ the solute behaves as a
``structure breaker'', and interestingly, its effect on the $T_{MD}$
is also more significant as the size of the R-site increases, to the
point that the drop of the TMD for the largest concentration
considered is three times larger for the heteronuclear
model. Unfortunately, we are not aware of any experimental
investigation of the pressure dependence of $\Delta
T_{MD}(x_{ROH})$, but since the net effect of pressure is to reduce
the range of  anomalous behavior (in real fluid by breaking the hydrogen bond
network, in our model by displacing particles towards to first range
of the potential), that fact that the effect is maximized when the
volume of the solute is larger is understandable from a enthalpic
point of view. 

\begin{figure}[!htb]
  \includegraphics[clip,scale=.7]{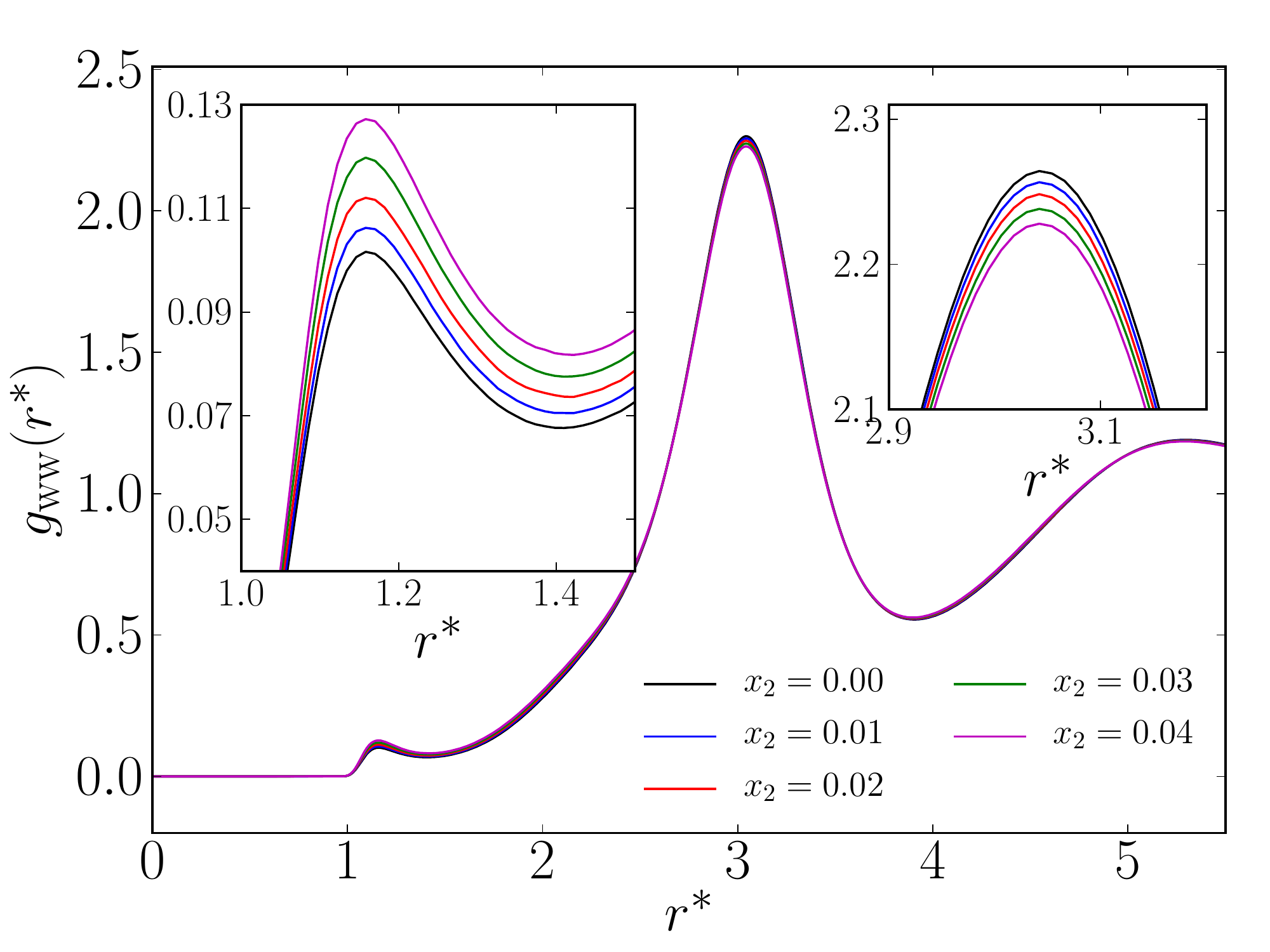}

    \includegraphics[clip,scale=.7]{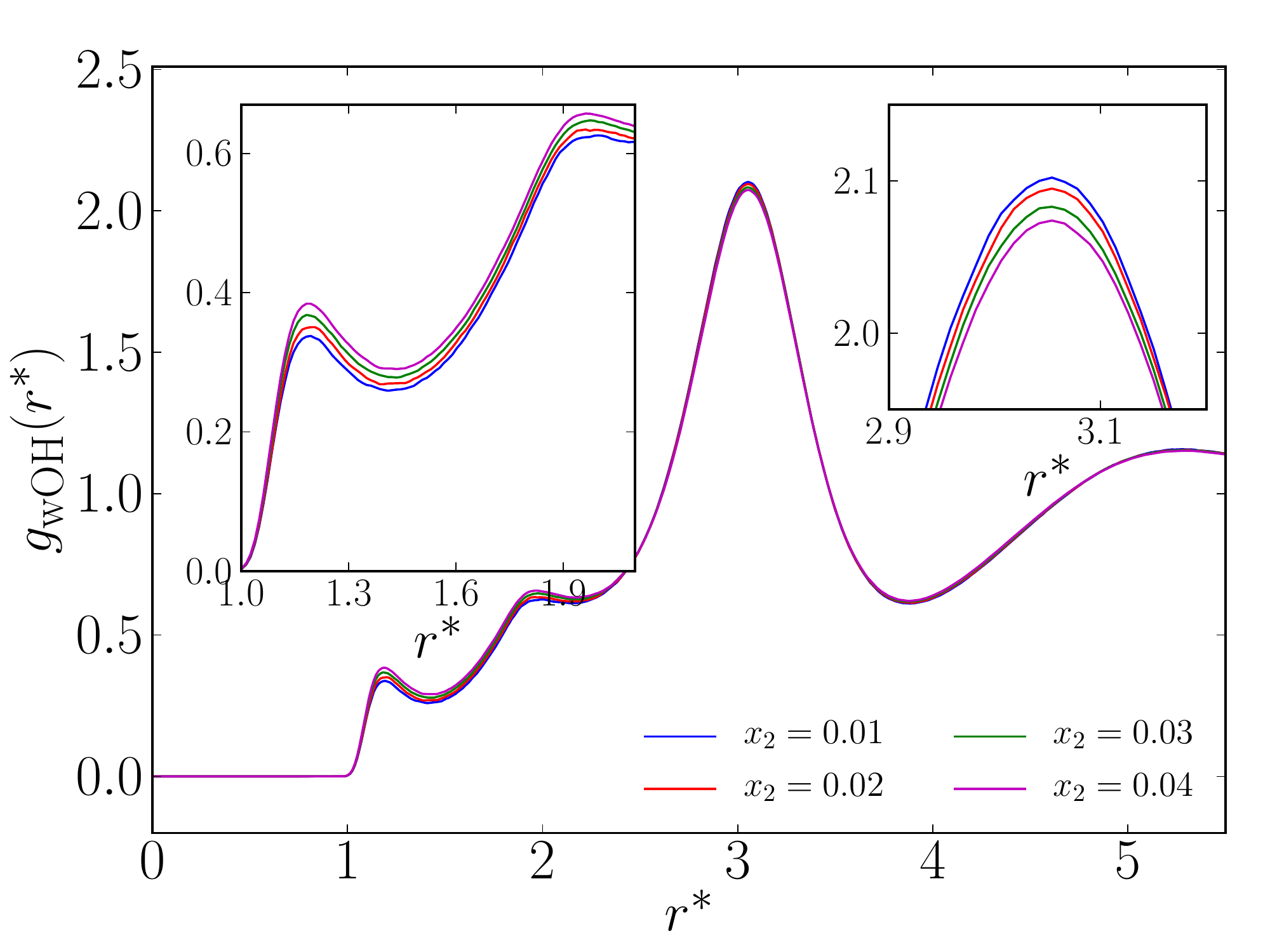}
  \caption{Water-water (up) and Water-OH (down) radial distribution
    functions for $P^*=6.8$ and
    $T^*=0.4$ for various solute concentrations. The insets show  zoom
    of the regions around the first two maxima.}
  \label{fig:gr_compare}
\end{figure}

From a microscopic point of view,  structural effects of the addition
of solute should be visible in the water-water and water-OH  pair distribution
functions. These are plotted in the
Figures~\ref{fig:gr_compare} for $P^*=6.8$ and $T^*=0.4$. One observes that
a small number water particles move into the first scale of the
potential (more compact structures), but at the same time, the area corresponding to the second
repulsive range of the potential ($r\approx 2\sigma$) becomes more
populated, which is particularly visible in the evolution of the
second maximum of the $g_{wOH}$ site-site
function. In this way, the addition of solute molecules leads to an
increase of open structures and more compact ones. The balance between
these open and compact structures is correlated with the subtle change
from $\Delta T_{MD} >0$ to $\Delta T_{MD} <0$ as $x_{ROH}$ grows. 

\subsection{Excess thermodynamic properties}
Excess thermodynamic properties of a mixture are defined as the difference
between the values of a given thermodynamic quantity and those that would
be obtained in an ideal mixture. For a given quantity, $A$, the
corresponding excess property is defined by
\begin{equation}
  \label{eq:exc_def}
  A^E(x_2,p,T,)=A(x_2,p,T)-\cch{x_2A^0_2(p,T) + (1-x_2)A^0_1(p,T)}
\end{equation}
where $A(x_2,p,T)$ is the value $A$ in binary mixture of a given
composition defined by mole fractions $(x_1,x_2)$.  $A^0_1$ and $A^0_2$ are
the values of  $A$ for the pure components at the same thermodynamic
state. Quantities of interest in binary mixtures are the excess volume $V^E$,
enthalpy, $H^E$, and specific heat at constant pressure
$c_P^E$. Excess entropy is also of interest, but it is not directly
accessible in MD calculation.  Excess volumes, $V^E$, are determined from the
average volume values obtained along NPT simulations for the mixtures
and pure components. Similarly, excess enthalpy is obtained from 
the usual expression
\begin{equation}
  H^E(x_2,p,T)=U^E\prt{x_2,p,T}+PV^E\prt{x_2,p,T},
\end{equation}
and the excess internal energy, $U^E$, is also directly evaluated from
the MD runs for the mixture and pure components. The fluctuation of
the enthalpy provides a direct path for the calculation of the
specific heat at constant  
pressure, $c_p^E$,
\begin{equation}
  \label{eq:exc_enth}
  c_p(x_2,p,T)=\prt{\frac{\partial H(x_2,p,T)}{\partial T}}_P
  \simeq \aver{\prt{H\prt{x_2,p,T}-\aver{H\prt{x_2,p,T}}}^2}_{NpT}.
\end{equation}
and therefore, 
\begin{equation}
  \label{eq:exc_cp}
  c_p^E(x_2,p,T)=c_P(x_2,p,T)-
  \cch{ x_2c^0_{p,2}(p,T)+(1-x_2)c^0_{p,1}(p,T)}.
\end{equation}
This property requires extremely long simulation runs, and we have
assessed the validity of our results comparing the results of the
fluctuation approach to those obtained by numerical differentiation of the
enthalpy with respect to temperature, for specific points. 

\begin{figure}[!htb]
   \includegraphics[scale=.4]{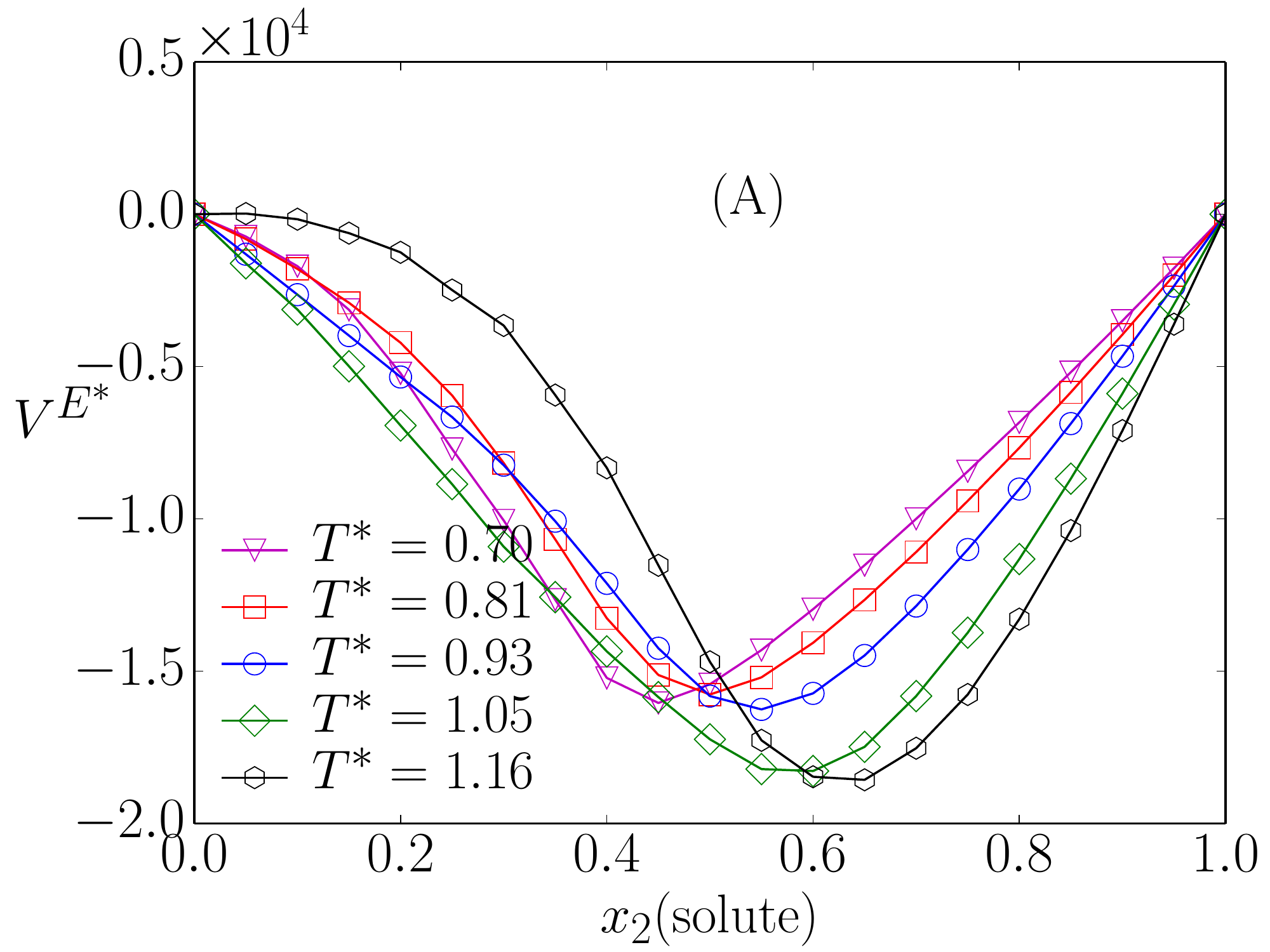}
   \includegraphics[scale=.4]{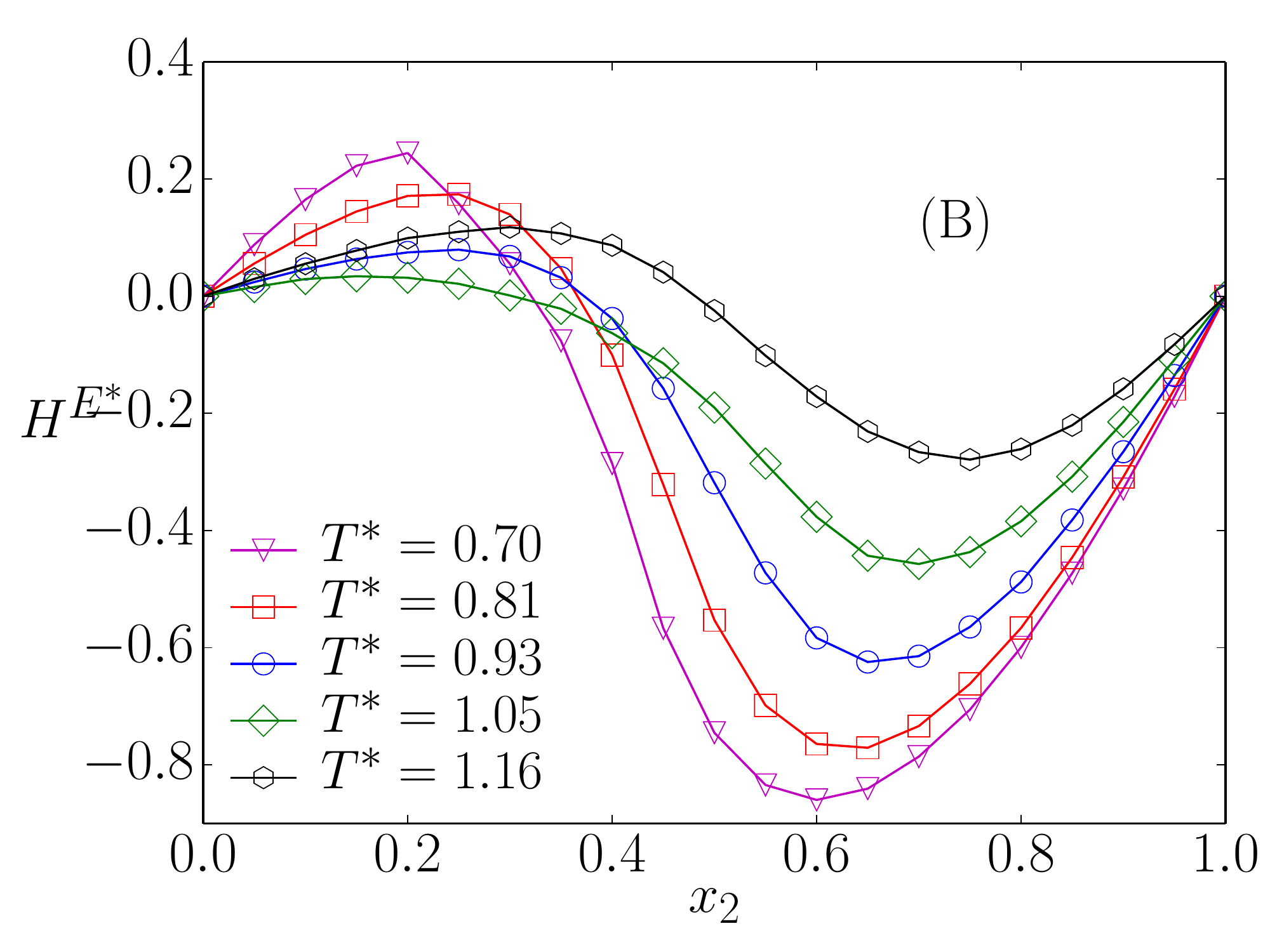}
   \includegraphics[scale=.4]{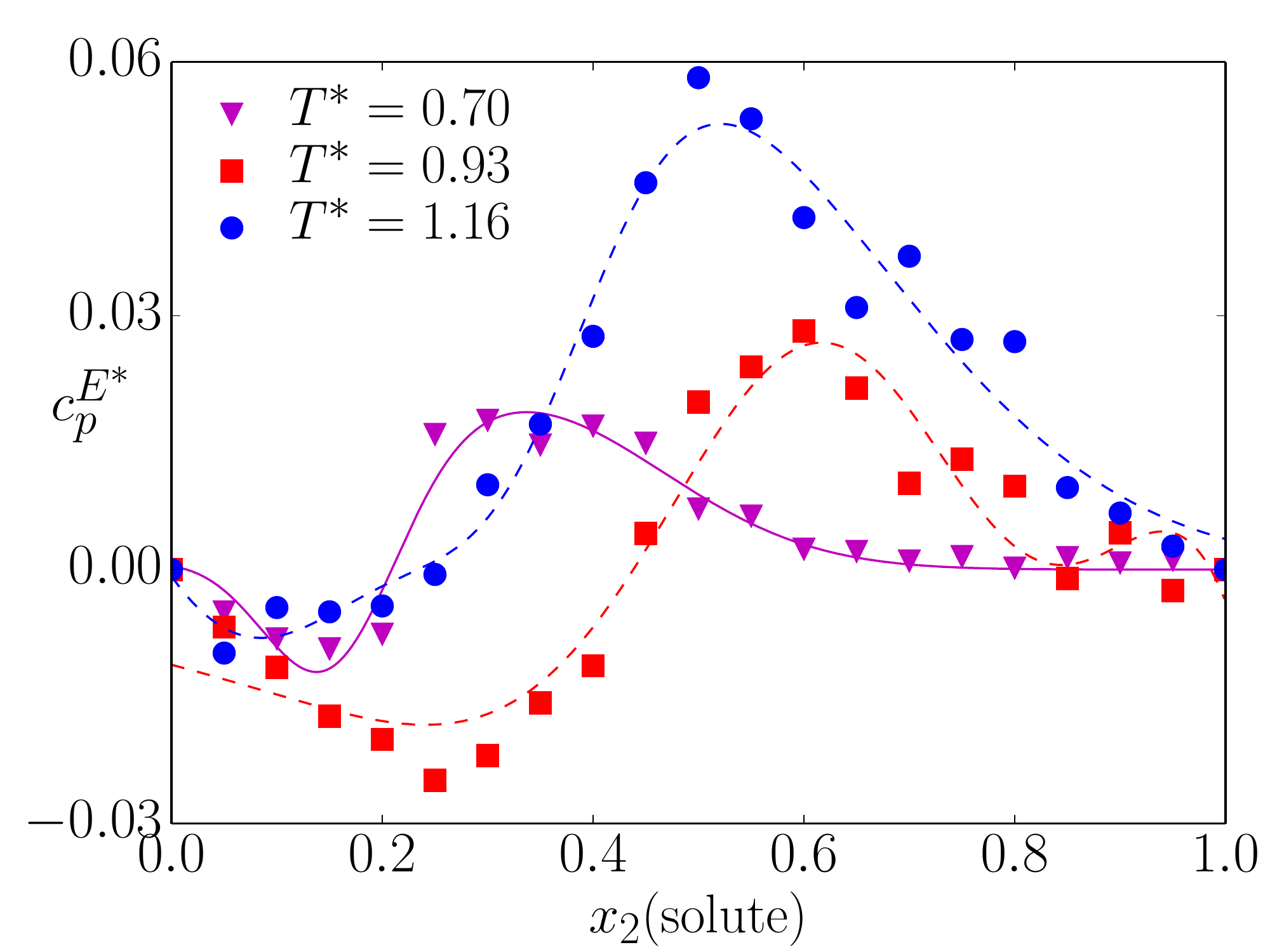}
   \caption{Excess thermodynamics of our water-ROH mixture model
     (heteronuclear model).  (A) Excess volume; (B) excess enthalpy;
   (C) excess specific heat. In all figures the symbols are 
   data obtained from MD simulations.  In 
   graphs (A) and (B) lines are drawn as a guide to the eyes. In graph 
   (C), to compensate the dispersion of the simulated date, the curve
   is a least squares fit. All calculations were done at pressure $P^*= 18$.}
   \label{fig:excs}
\end{figure}

Our results for the excess thermodynamics of our mixture system
(heteronuclear model) are collected  in  the Figure~\ref{fig:excs}.
The excess volume exhibits the typical volume contraction of the
mixture, characteristic of short chain
alcohols~\cite{Subramanian2013,Mijakovic2014,Gonzalez-Salgado2016}. This
is in agreement with the observed behavior in $g(r)$
(Figure~\ref{fig:gr_compare}), in which is seen that water particles
move closer to each other when solute is incorporated. 

The situation is somewhat different for the excess enthalpy. Here our
model exhibits a minimum for alcohol-rich solutions, in contrast with
the experimental situation for methanol~\cite{Gonzalez-Salgado2016},
ethanol~\cite{Mijakovic2014} and tert-butanol~\cite{Subramanian2013}. In
these cases the minimum occurs for water-rich
conditions. Moreover, tert-butanol~\cite{Subramanian2013} 
excess enthalpies change
sign as the concentration of alcohol increases but, contrary to our
model's behavior, positive values occur at high alcohol
concentrations. As shown by Gonz\'alez-Salgado and
coworkers~\cite{Gonzalez-Salgado2016} these discrepancies could be cured
by a simple tuning of the cross interaction parameters. Even with more
or less sophisticated models for the pure alcohol and water, excess
properties can be even qualitatively wrong when Lorentz-Berthelot
mixing rules are used~\cite{Mijakovic2014,Gonzalez-Salgado2016}.

Finally, in the lower graph of tne Figure~\ref{fig:excs} we have the excess
constant pressure heat capacity, as obtained from
Eq.~\ref{eq:exc_cp}. The model performance for  the excess heat capacity is
correlated with that of the excess enthalpy. Again here we observe the
presence of a maximum in agreement with experimental results for
methanol~\cite{Gonzalez-Salgado2016} and
tert-butanol~\cite{Subramanian2013}, but  the model
predicts its position at somewhat higher concentrations of
alcohol. Nonetheless, we can say that at relatively low temperature
the increase of the heat capacity reflects the structure-making
character of our solvent, in accordance with the experimental
findings.  Again, discrepancies such as the presence of negative
values of the excess heat capacity or the shift of the maxima to
regions of higher alcohol concentration can be tuned by a careful
choice of the cross interaction parameters.

\section{Conclusions}\label{sec:concl}

In summary, we have presented a detailed computer simulation study of
a simple model for diluted alcohol-water mixtures, in which the
interactions involving hydrogen bonding are represented by a two-scale
potential which is known to reproduce a good number of water
anomalies. Our results for the dependence of the temperature of
maximum density on the solute concentration, are in qualitative
agreement with the experimental behavior of methanol and t-butanol
solutions, whose molecules are modeled by a homonuclear and heteronuclear dimer
respectively. These results indicate that for a small range of
concentrations and up to certain values of pressure, these
hydrogen-bonding-like solutes tend to enhance the open structures of
water and hence increase the TMD, behaving as ``structure-makers''. 

As pressure increases the ``structure-breaker''
character of the solutes is enhanced, being larger as the size of the
alkyl group grows. This is understandable as the presence of the
apolar group as pressure increases makes more unfavorable the open
structures which are responsible for the anomalous behavior of the
model. This enthalpic effect increases with the size of the solute
molecule. 

 Future work will focus on the dynamic anomalies (e.g. the
increase of the diffusion constant with pressure) which are known
to be influenced in a similar fashion when diluted hydrogen bonded
solutes are present, in marked contrast with the effect of other
solutes, either polar or apolar. 

\acknowledgments
The authors gratefully acknowledge support from the Programa Ci\^encia
sem Fronteiras do Governo Federal de Brasil, for a Pesquisador
Visitante Especial Grant no. 401036/2014-6. 
EL  acknowledges the support from the Direcci\'on
General de Investigaci\'on Cient\'{\i}fica  y T\'ecnica under Grant
No. FIS2013-47350-C5-4-R.

\newpage
\bibliography{references.bib}

\end{document}